# On the N-dimensional hydrogen atom in momentum representation

## M. Hage-Hassan
Université Libanaise, Faculté des Sciences Section (1)
Hadath-Beyrouth Liban

## Abstract

We derive the wave function for N-dimensional hydrogen atom in the momentum representation with the phase factor using the generating function method and Hankel's integral.

### 1-Introduction

The problem of the hydrogen atom has played a central role in the development of Quantum mechanics. The wave function of the hydrogen atom in momentum space has Solved by Fock's stereographic projection method [1-6]. Recently, The Fock method has been generalized for the determination of the wave function in N-dimensions space and was applied by several authors at different physical problems [7-8]. But the Fock method is not a direct calculation of the Fourier transform of the wave function of coordinates and further the phase factors is can not be determined [7-8] correctly.

In a previous paper [9-10] we presented a new and elementary method for the determination of the wave function in momentum space for two and three dimensions using the generating function and Hurwitz transformations for a direct integration of the Fourier transform. But Hurwitz transformations are valid only for N = 2, 3, 5 and 9.

In this work we present, the generalization for N≥3 by using the technique of generating function and the Hankel's integral of Bessel functions and therefore we determine the wave function in momentum space with the exact phase factor for any order N.
We also show that the phase factors of wave functions representing the coordinates and momentum spaces are dependent.

The paper is structured as follows. First in the second part we outline the derivation of the wave function of coordinates. We determine the wave function in momentum space in the third part. The fourth part is devoted to a comment on the phase's factors.

### 2. The N-dimensional hydrogen atom in representation space

In this part we exhibit only the well known wave function solution of the Schrödinger equation for the hydrogen atoms in N-dimensions [13-15].

## 2.1 The wave function of hydrogen atom in representation space

The Schrödinger equation of the hydrogen atom in N-dimensions space [8] is

$$\left(-\frac{\hbar^2}{2\mu}\Delta_N - \frac{e^2}{r}\right)\Psi(\vec{r}) = E\Psi(\vec{r}) \qquad (2.1)$$

Where $\mu$ is the reduced mass.

We write $\vec{r} = (x_1, x_2, \ldots, x_N)$ in spherical coordinates as $\vec{r} = (r, \theta_1, \theta_2, \ldots, \theta_{N-2}, \varphi)$.

$$\begin{aligned}
x_1 &= r\sin\theta_1 \sin\theta_1 \sin\theta_2 \ldots \sin\theta_{N-2} \cos\varphi \\
x_2 &= r\sin\theta_1 \sin\theta_2 \sin\theta_3 \ldots \sin\theta_{N-2} \sin\varphi \\
x_3 &= r\sin\theta_1 \sin\theta_2 \sin\theta_3 \ldots \cos\theta_{N-2} \\
&\vdots \\
x_{N-1} &= r\sin\theta_1 \cos\theta_2 \\
x_N &= r\cos\theta_1
\end{aligned} \qquad (2.2)$$

With $0 \leq \theta_j \leq \pi$, $0 \leq \theta_j \leq \pi$, $j = 1, \ldots, N-2$, $0 \leq \varphi \leq 2\pi$.

And atomic unit are used through the text.

The method of separation of variables is used by many authors [13-15] for the resolution of the Schrödinger equation and we write only the solution:

$$\Psi_{n,l,\{\mu\}}(\vec{r}) = R_{n,l}(r)Y_{l,\{\mu\}}(\Omega_r) = N_{n,l}\,\omega^{N/2}(\omega r)^l\, e^{-\frac{r}{2\lambda}} L_{n-l-1}^{2l+N-2}(\omega r) Y_{l,\{\mu\}}(\Omega_r) \qquad (2.3)$$

$$N_{n,l} = \left\{\frac{(n-l-1)!}{2(n+(N-3)/2)(n+l+N-3)!}\right\}^{1/2}, \quad \omega = 2\delta_n = \frac{2}{(n+(N-3)/2)}$$

## 2.2 The radial function

$L_{n-l-1}^{(\alpha)}(x)$ Are the Laguerre polynomials

$$\int_0^\infty e^{-x} x^\alpha (L_n^\alpha(x))^2\, dx = \frac{\Gamma(\alpha+n+1)}{n!} \qquad (2.4)$$

The generating function of Laguerre polynomials [16-17] is:

$$\sum_{n=0}^\infty z^n L_n^{(\alpha)}(x) = \frac{1}{(1-z)^{\alpha+1}} e^{-\frac{z}{1-z}x} \qquad (2.5)$$

$$\frac{d^k}{dx^k} L_n^{(\alpha)}(x) = (-1)^k L_{n-k}^{(\alpha+k)}(x)$$

$$\alpha + k = 2l + N - 2,\ n - k = n - l - 1$$

Consequently $\alpha = l + N - 3$

We deduce also that

$$\frac{(-z)^{l+1}}{(1-z)^{2l+N}} \exp[-x\frac{z}{(1-z)}] = \sum_{n=0}^\infty z^n L_{n-l-1}^{2l+N-2}(x) \qquad (2.6)$$

## 2.3 The angular function

$Y_{l,\{\mu\}}(\Omega_r)$ Is the Hyperspherical function

$$Y_{l,\{\mu\}}(\Omega_r) = \frac{1}{\sqrt{2\pi}} A_{l,\{\mu\}} e^{im\varphi} \prod_{j=1}^{N-2} C_{\mu_j-\mu_{j+1}}^{\alpha_j+\mu_{j+1}}(\cos\theta_j)(\sin\theta_j)^{\mu_{j+1}} \qquad (2.7)$$

With
$$A_{l,\{\mu\}} = \prod_{j=1}^{N-2} \Gamma(\alpha_j + \alpha_{j+1}) \sqrt{\frac{(\alpha_j + \mu_j)(\mu_j - \mu_{j+1})!}{\pi 2^{1-2\alpha_j - 2\mu_{j+1}} \Gamma(2\alpha_j + \mu_j + \mu_{j+1})}}$$

And $2\alpha_j = N - j - 1, \, l = \mu_1 \geq \mu_2 \geq \ldots \geq |\mu_{N-1}| = |m|, \, (l,\{\mu\}) = (l, \mu_2, \ldots, \mu_{N-1})$.

$C_n^\alpha(\cos\theta)$ Is the Gegenbauer polynomial of degree n and parameter α.

### 3. The N-dimensional hydrogen atom in the momentum representation

Using the development of the free wave in space of N-dimensions, the generating functions of Laguerre polynomials and Hankel's integral we determine the generating function of momentum representation and hence the wave function in momentum space.

**3.1 The Generating function and the momentum representation**

The wave function of hydrogen atom in momentum representation is:

$$\Psi_{n,l,\{\mu\}}(\vec{p}) = \frac{1}{(2\pi)^{N/2}} \int e^{-i\vec{p}\cdot\vec{r}} \Psi_{n,l,\{\mu\}}(\vec{r}) d\vec{r} \tag{3.1}$$

$$= \frac{N_{n,l}}{(2\pi)^{N/2}} \int e^{-i\vec{p}\cdot\vec{r} - \frac{\omega r}{2}} (\omega r)^l L_{n-l-1}^{2l+N-2}(\omega r) Y_{l,\{\mu\}}(\Omega_r) d\vec{r} \tag{3.2}$$

$d\vec{r}$ Is defined by:
$$d\vec{r} = dx_1 dx_2 \ldots dx_N = r^{N-1} d\Omega_r. \tag{3.3}$$

We note that $i\vec{p}\cdot\vec{r} + \delta r$ can be regarded as the scalar product of two vectors in Euclidean space $E_{N+1}$. The first one is a vector of zero lengths $(x_1, x_2, \ldots, x_N, ir)$ and the second vector is defined by the angles $\vec{K} = (\theta, \theta_1^k, \ldots, \theta_{N-1}^k)$ and the length $K = \sqrt{p^2 + \delta_n^2}$

Using the development of the wave function of the free particle in N-dimensions [14-15]:

$$e^{i\vec{p}\cdot\vec{r}} = (2\pi)^{N/2} \sum_{[L]} i^l Y_{[L]}(\Omega_r) Y_{[L]}^*(\Omega_p) J_\nu(pr)/(pr)^{\frac{N}{2}-1}, \quad \nu = l + \frac{N}{2} - 1 \tag{3.4}$$

We find

$$\Psi_{n,l,\{\mu\}}(\vec{p}) = \frac{1}{(2\pi)^{N/2}} \int e^{-i\vec{p}\cdot\vec{r}} \Psi_{n,l,\{\mu\}}(\vec{r}) d\vec{r}$$

$$= N_{n,l} \left( \int_0^\infty (\omega r)^l e^{-\frac{\omega r}{2}} L_{n-l-1}^{2l+N-2}(\omega r) J_\nu(pr)/(pr)^{\frac{N}{2}-1} r^{N-1} dr \right) \left( (-i)^l Y_{[L]}(\Omega_p) \right) \tag{3.5}$$

Multiply by $1/(N_{n,l}) z^n /(\omega^{l+N/2})$ and do the summation we write first the generating function for the basis and the generating function with ω=constant which is very useful for the determination of wave function in momentum space.

$$G(p, z, \delta_n) = \sum_{n=0}^\infty \frac{(p)^{\frac{N}{2}-1}}{N_{n,l}} \frac{z^n}{\omega^{l+N/2}} \Psi_{n,l}(\vec{p}) \tag{3.6}$$

And
$$G(p, z, \delta) = \sum_{n=0}^\infty z^n \left( \int_0^\infty (r)^l e^{-\frac{r}{2\lambda}} L_{n-l-1}^{2l+N-2}(\omega r) J_\nu(pr)/(pr)^{\frac{N}{2}-1} r^{N-1} dr \right)$$

$$= \frac{(-z)^{l+1}}{(1-z)^{2l+N-1}} \int_0^\infty e^{-\gamma r} J_\nu(pr) r^{\nu+1} dr \qquad \nu = l + \frac{N}{2} - 1 \qquad (3.7)$$

With
$$\gamma = \frac{\omega}{2}\frac{1+z}{1-z}, \omega = 2\delta \qquad (3.8)$$

**3.2 Derivation of the generating function**

Using Hankel's integral [17]:
$$\int_0^\infty e^{-\gamma r} J_\nu(pr) r^{\nu+1} dr = \frac{2\gamma(2p)^\nu \Gamma(\nu+3/2)}{\sqrt{\pi} \times (\gamma^2 + p^2)^{\nu+3/2}}, [\operatorname{Re}(\nu) > -1] \qquad (3.9)$$

We find the generating function as follows:
$$\frac{(-z)^{l+1}}{(1-z)^{2l+N-1}} \frac{2\gamma(2p)^\nu \Gamma(\nu+3/2)}{\sqrt{\pi} \times (p^2+\gamma^2)^{\nu+3/2}} =$$

$$= \frac{\omega}{\sqrt{\pi}} \frac{1+z}{(1-z)^{2l+N}} (2p)^{l+\frac{N}{2}-1} \Gamma(l+\frac{N+1}{2}) \frac{(-z)^{l+1}(1-z)^{2l+N+1}}{(p^2(1-z)^2 + \delta^2(1+z)^2)^{l+\frac{N+1}{2}}}$$

$$= \frac{2\delta(2p)^{l+\frac{N}{2}-1} \Gamma(l+\frac{N+1}{2})}{\sqrt{\pi}(p^2+\delta^2)^{l+\frac{N+1}{2}}} \frac{(-z)^{l+1}(1-z^2)}{(1-2zx+z^2)^{l+\frac{N+1}{2}}}, x = \left(\frac{p^2-\delta^2}{p^2+\delta^2}\right) \qquad (3.10)$$

This function is the generalization of the generating function (4.10) of the paper [9], by a minus sign due to the derivation of (4.9).

**3.3 The Wave function in the momentum representation**

We have [9] 
$$\frac{1}{n!}\frac{d^n}{dz^n} G(p,z,\delta_n)\bigg|_{z=0} = \frac{1}{n!}\frac{d^n}{dz^n}[G(p,z,\delta)|\delta = \delta_n]\bigg|_{z=0} \qquad (3.11)$$

Using the development
$$\frac{(-z)^{l+1}(1-z^2)}{(1-2zx+z^2)^{l+\frac{N+1}{2}}} = (-1)^{l+1} \sum_{n=1}^\infty z^n [C_{n-l-1}^{l+\frac{N+1}{2}}(x) - C_{n-l-3}^{l+\frac{N+1}{2}}(x)] \qquad (3.12)$$

And the formula [16] $\qquad (n+\alpha)C_{n+1}^{(\alpha-1)}(x) = (\alpha-1)[C_{n+1}^{(\alpha)}(x) - C_{n-1}^{(\alpha)}(x)] \qquad (3.13)$

We find that the expression (3.11) may be written as

$$\frac{(p)^{\frac{N}{2}-1}}{N_{n,l}(2\delta_n)^{l+\frac{N}{2}}} \Psi_{n,l}(\vec{p}) = (-1)^{l+1} \frac{(2\delta_n)(2p)^{l+\frac{N}{2}-1} \Gamma(l+\frac{N-1}{2})(n+\frac{(N-3)}{2})}{\sqrt{\pi}(p^2+\delta_n^2)^{l+\frac{N+1}{2}}} C_{n-l-1}^{l+(N-1)/2}(x)$$

Finally we derive the wave function in the momentum representation

$$\Psi_{n,l,\{\mu\}}(\vec{p}) = -(i)^l \left\{\frac{(n-l-1)!(n+(N-3)/2)}{2\pi(n+l+N-3)!}\right\}^{1/2} \frac{2^{2l+N}(\delta_n)^{\frac{N}{2}+1}(\delta_n p)^l \Gamma(l+\frac{N-1}{2})!}{(p^2+\delta_n^2)^{l+\frac{N+1}{2}}} \times$$

$$C_{n-l-1}^{l+(N-1)/2}(x) Y_{[L]}(\Omega_p) \qquad (3.14)$$

This hyperphysical function may be written in term of the components of the vector defined by the angles $\vec{K}' = (-\frac{\pi}{2} + 2\theta, \theta_1^k, \ldots, \theta_{n-1}^k)$ and the length $K' = \sqrt{\vec{P}^2 + \delta_n^2}$.
This vector may be derived from $\vec{K}$ by rotation about the vector perpendicular to the space $E_N$ with angle of rotation $-\frac{\pi}{2} + \theta$.
We can also determine the representation {p} by this method if the potential has an additional term $A/r^2$.

## 4. Appendix: a remark on the phase factor

We show by an example that inattention error are repeated in many textbooks of quantum mechanics [18-20] and work [7-8] on the phase factor of the basis of the representation {p} of the harmonic oscillator and the hydrogen atom.
In quantum mechanics the Schrödinger equation of harmonic oscillator is

$$H\psi(x) = E\psi(x) \quad (4.1)$$

$$H = \frac{1}{2m}(p_x^2 + m^2\omega^2 x^2) \quad [x, p_x] = i\hbar \quad (4.2)$$

**4.1** In the representation {x} we find

$$\psi_n(x) = \left(\frac{\sqrt{m\omega}}{\sqrt{\pi\hbar}\, 2^n n!}\right)^{+\frac{1}{2}} e^{-\frac{m\omega x^2}{2\hbar}} H_n\left(\sqrt{\frac{m\omega}{\hbar}} x\right) \quad (4.3)$$

The well known generating function is $G(t, x, \psi) = \sum \varphi(t)\psi_n(x)$

With $\varphi(t) = \frac{t^n}{\sqrt{n!}}$ is the basis of Fock-Bargmann space with measure

$$d\mu(t) = \frac{1}{\pi} e^{-t\bar{t}} du dv, t = u + iv$$

**4.2** We deduce the representation {p} by putting $x = i\hbar d/dp_x$ in Schrödinger equation and we find the solution:

$$\psi'_n(p) = \left(\frac{1}{\sqrt{\pi m\omega\hbar}\, 2^n n!}\right)^{+\frac{1}{2}} e^{-\frac{1}{2m\omega\hbar}\frac{p^2}{2}} H_n\left(\sqrt{\frac{1}{m\omega\hbar}} p\right) \quad (4.4)$$

**4.2.1** The phase factor is can not be determined in this case it is chosen equal to one.
**4.2.2** the direct calculation of the representation {p} by the Fourier transformation
Using the generating function gives [21] $\psi_n(p) = (-i)^n \psi'_n(p)$.

**4.3** If we perform the calculations of $\langle x | p_x \rangle$ using $\psi'_n(p)$:

$$\langle x | p_x \rangle = \sum_{n=0}^{\infty} \overline{\psi}_n(x)\psi'_n(p) = \int_0^{\infty} G(\bar{t}, x, \psi)G(t, p, \psi')d\mu(t)$$

We find $\quad \langle x | p_x \rangle = \delta(X - P), X = \sqrt{m\omega}$ and $P = \sqrt{\frac{1}{m\omega}} \quad (4.5)$

This result is incorrect.

**4.4** But if we perform the calculation with $\psi_n(p)$

$$\langle x | p_x \rangle = \sum_{n=0}^{\infty} \overline{\psi}_n(x)\psi_n(p) = \int_0^{\infty} G(\bar{t},x,\psi)G(t,p,\psi)d\mu(t)$$

We obtain the correct expression $\langle x | p_x \rangle = \dfrac{1}{\sqrt{2\pi\hbar}} e^{i(x p_x)/\hbar}$. (4.6)

So we deduce that we are not free to choice the phase factors of representations {x} and {p} in an independent manner.